\documentclass[10pt]{article}
\PassOptionsToPackage{style=numeric-comp}{biblatex}
\usepackage[letterpaper]{geometry}
\usepackage{hicss}
\usepackage{times}
\usepackage[none]{hyphenat}
\usepackage{url}
\usepackage{latexsym}
\usepackage{indentfirst}
\usepackage{graphicx}
\usepackage{xcolor}
\usepackage{amsmath}
\usepackage{tabularx}
\usepackage{float}
\usepackage{enumitem}
\usepackage{dblfloatfix}
\graphicspath{{images/}}
  \usepackage[
    style=ieee,
  ]{biblatex}
\newif\ifpreprint
\preprinttrue
\ifpreprint
  \usepackage{fancyhdr}
\fi
\newcommand{\uti}[1]{u_{#1}}

\newcommand{\RF}[0]{R_\mathrm{F}}
\newcommand{\VF}[1]{V_\mathrm{F{#1}}}

\newcommand{\RT}[0]{R_\mathrm{T}}

\newcommand{\RD}[0]{R_\mathrm{D}}

\newcommand{\VTl}[0]{V_\mathrm{T0}}
\newcommand{\VDl}[0]{V_\mathrm{D0}}

\newcommand{\VPFCin}[0]{V_\mathrm{in}}
\newcommand{\IPFCin}[0]{I_\mathrm{in}}

\newcommand{\Vrec}[0]{V_\mathrm{rec}}
\newcommand{\Irec}[0]{I_\mathrm{rec}}

\newcommand{\Vbrloss}[0]{V_\mathrm{br}^\mathrm{l}}
\newcommand{\Vbloss}[0]{V_\mathrm{b}^\mathrm{l}}
\newcommand{\Ibswloss}[0]{I_\mathrm{b,sw}^\mathrm{l}}

\newcommand{\Vbin}[0]{V_\mathrm{b,in}}
\newcommand{\Vbout}[0]{V_\mathrm{b,out}}

\newcommand{\Illcswloss}[0]{I_\mathrm{llc,sw}^\mathrm{l}}

\newcommand{\fsw}[1]{f_\mathrm{sw}^\mathrm{#1}}

\newcommand{\ton}[0]{t_\mathrm{on}}
\newcommand{\toff}[0]{t_\mathrm{off}}

\newcommand{\Uclus}[1]{u_{#1}}
\newcommand{\Userv}[2]{u_{#1#2}}

\begin{document}
\ifpreprint
  \pagestyle{fancy}
  \thispagestyle{fancy}
  \fancyhf{}
  \renewcommand{\headrulewidth}{0pt}
  \renewcommand{\footrulewidth}{0.4pt}
  \fancyfoot[C]{\footnotesize\itshape Preprint: Accepted for publication at the Hawaii International Conference on System Sciences (HICSS-60).}
\fi

\title{Steady-State Equivalent Circuit Model for Data Center Loads\\
}
\author{
Muhammad Hamza Ali$^{1}$, Peng Sang$^{1}$, Hyeon Woo$^{2}$, Hyein Kang$^{2}$, Sungyun Choi$^{2}$, Amritanshu Pandey$^{1}$ \\
$^{1}$University of Vermont, Burlington, VT, USA \\
$^{2}$Korea University, Seoul, Republic of Korea \\
\texttt{\{muhammad-hamza.ali, peng.sang, amritanshu.pandey\}@uvm.edu} \\
\texttt{\{wh9714, hyeinkang\}@korea.ac.kr, sungyun@korea.ac.kr} \\
}

\maketitle
\vspace{-8pt}
\begin{abstract}
\noindent Planners currently represent data centers as aggregate constant-PQ or ZIP loads in steady-state interconnection and contingency studies.
These aggregate models are computationally convenient.
However, they obscure the electrical relationship between computational workloads, server utilization, and grid-side demand.
They ignore the internal power-electronic conversion stages of IT loads and assume homogeneous workload distributions across the compute clusters.
This hides operating-point-dependent converter losses and efficiency variations.
We propose a steady-state \textit{equivalent-circuit model} (ECM) for data centers, which explicitly builds circuit models for IT loads, power supply units, cooling, and auxiliary systems.
For power supply units, the equivalent circuit model explicitly represents internal power-electronic conversion stages.
For IT loads, we develop a utilization-dependent server power model, and we combine it with loss-aware ECMs of power supply units. This approach captures the grid-side impact of heterogeneous workload distributions while preserving compatibility with conventional power-flow analysis.
We evaluate this data center ECM in large-scale transmission power flows, using Monte Carlo simulations under heterogeneous and homogeneous cluster utilization. In comparison with the fixed-efficiency constant-PQ model, the ECM predicts that the most stressed line exceeds its thermal limit in about 30\% of Monte Carlo samples. The results further show that homogeneous server utilization overstates line-loading variability by 17\%-46\% relative to heterogeneous server utilization, depending on the intra-cluster workload correlation.

\end{abstract}
\vspace{10pt}
\noindent \textbf{Keywords:} Data center, equivalent circuit model, PSU conversion losses, large-load interconnection, server utilization.
\section{Introduction}
\noindent \textbf{Motivation:} Data centers are becoming key components of modern power systems and, in some regions, contribute significant demand.
As of today, Dominion Power reports that around 28\% of the U.S. state of Virginia's electricity sales are from data centers in 2025~\cite{dominion2025annualreport}.
Today, grid planners model data centers as aggregate constant-PQ or ZIP loads in steady-state power-flow studies.
These models are computationally convenient, easy to parameterize, and can be used to run thousands of power flow studies quickly.
However, these aggregate models obscure internal operating states of data centers.
They fail to capture how grid-side behavior depends on workload allocation and power electronic interfaces.
A recent NERC incident review provides a concrete example.
During a July 2024 event in the Eastern Interconnection in Virginia, a 230-kV transmission line fault led to a customer-initiated simultaneous reduction of approximately 1{,}500~MW of voltage-sensitive load.
NERC identified this exclusively as data center load in an area with high data center concentration~\cite{nerc2025incident}.
Following these concerns, NERC and Dominion highlighted improved large-load interconnection requirements and modeling practices~\cite{parker2025dominion,NERC2025_LargeLoadsAlert,nerc2025incident}.

\begin{figure}[t]
    \centering
    \includegraphics[width=\columnwidth]{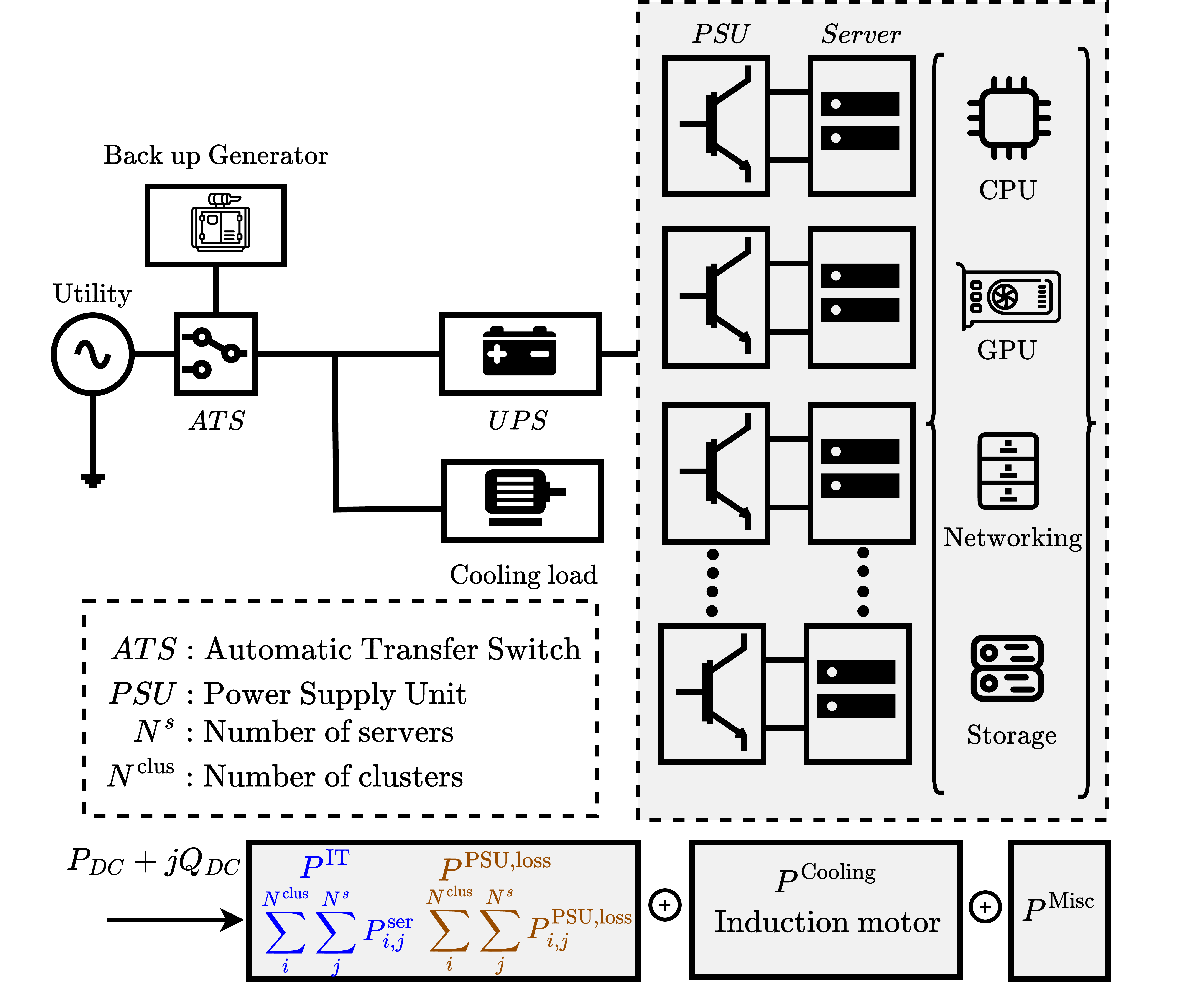}
    \caption{General electrical architecture of a data center, including utility supply, backup generation, UPS, cooling load, and server-level IT load components.}
    \label{fig:dc_architecture}
\end{figure}

Most transmission planning and contingency studies rely on steady-state models that can be evaluated repeatedly across many power-flow scenarios.
While constant-PQ and ZIP models satisfy this computational requirement, they collapse the data center into a single terminal load.
A data center actually consists of several major load categories.
These include IT load, cooling load, and auxiliary load, as illustrated in Fig.~\ref{fig:dc_architecture}.
Server utilization and power supply units (PSU) conversion losses drive the IT load.
The cooling load represents HVAC systems.
The auxiliary load includes lighting, networking support, and other facility services.
In this paper, we use an equivalent-circuit approach to model these components: PSU stages for the IT load, an induction-motor equivalent circuit for the cooling load, and an aggregate equivalent load for auxiliary demand.

Data centers execute diverse workloads, including AI training, inference serving, and high-performance computing, resulting in a wide range of operating modes.
Different server-level workload allocations shift converter operating points and change power-conversion losses, even for the same aggregate IT demand.
For example, an LLM training task is load-balanced.
Parallel training is synchronized between servers in a cluster, making it homogeneous~\cite{narayanan2021efficient}.
Agentic AI workloads are inherently heterogeneous~\cite{raj2026understandinganalyzingoptimizingagentic}.
These tasks arrive with diverse hardware requirements and produce uneven server utilization.
A single aggregate load model cannot explicitly represent how grid-side power demand depends on workload distribution across servers.
Server-level workload allocations remain indistinguishable at the terminal of an aggregate load model.
However, these allocations produce different internal converter loading conditions, resulting in significantly different voltage and demand responses during large deviations from an operating point. In this work, we consider both homogeneous and heterogeneous workload allocation among servers.

\noindent \textbf{State-of-the-art:} Recent studies have developed dynamic and electromagnetic-transient models of data centers to assess grid reliability impacts.
These studies examine data-center electrical architectures, converter interfaces, protection behavior, and stability risks from fast-ramping loads~\cite{ross2026emtp_datacenter,sun2022dynamic,jimenez2025data,lu2026dynamic,kwon2025operational}.
These models support transient stability and disturbance analysis.
However, they primarily characterize dynamic response.
Dynamic and EMT studies often represent the upstream grid with a slack source or Thevenin equivalent to focus on transient equipment responses at the point of interconnection and reduce computational burden. In contrast, this paper develops a steady-state data-center load model for large-scale transmission analysis, where planners must efficiently evaluate thousands of contingency and interconnection scenarios.

\begin{figure}[t!]
    \centering
    \includegraphics[width=\columnwidth]{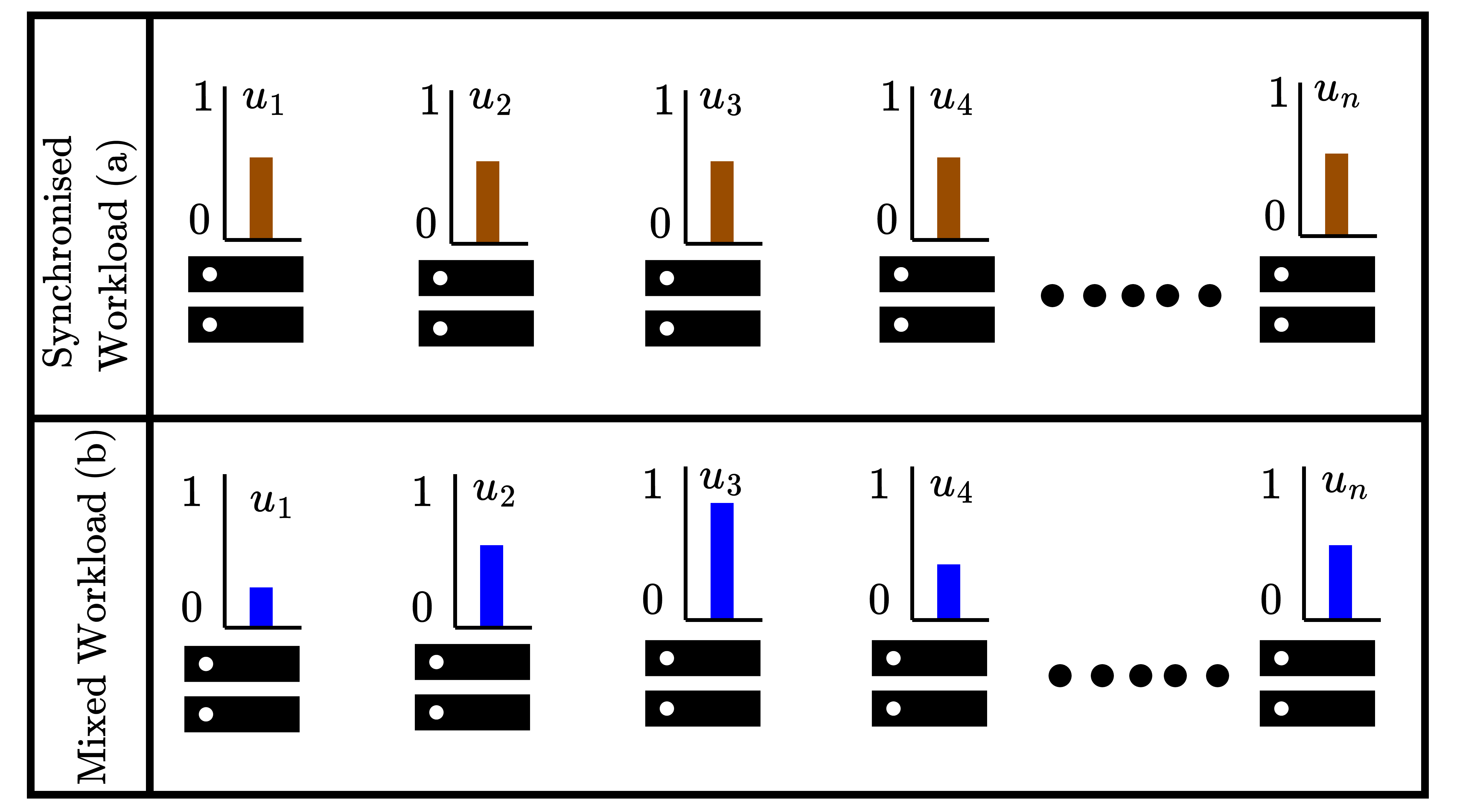}
    \caption{Server-level utilization $u \in [0,1]$ patterns under synchronized and mixed workloads. Synchronized workloads drive servers toward homogeneous utilization levels, while mixed workloads distribute utilization heterogeneously across servers.}
    \label{fig:server_utilization}
\end{figure}

On the other spectrum, a different body of literature focuses on data center energy usage and facility-level operation. 
These studies model interactions between IT equipment, cooling infrastructure, building energy performance, and facility-level power consumption~\cite{zhabelova2018comprehensive,rahmani2018complete,dayarathna2016data}.
They primarily support energy management and thermal analysis.
However, they are formulated to compute energy usage over time rather than instantaneous system-level power flow.
They do not provide a network-coupled electrical load model with a bus-level power interface.
They ignore voltage-dependent behavior and internal electrical operating states.
Aggregate facility demand under a large-deviation event such as a contingency cannot be inferred from aggregated load measurements, and future operating conditions are similarly hard to parameterize from nameplate capacity alone~\cite{fan2007power,barroso2018datacenter}. Workload characterization and cloud resource-management studies show that utilization varies across virtual machines, servers, applications, and allocation policies~\cite{khan2012workload,unuvar2014cloud,dong2023agent}.
Conversely, large-scale AI training workloads involve highly coordinated execution across large GPU clusters~\cite{narayanan2021efficient}.
Figure~\ref{fig:server_utilization} illustrates two representative server-level utilization patterns.
In the synchronized workload case, distributed AI training or tightly coupled HPC jobs drive many servers to similar utilization levels~\cite{dong2023agent}.
In the mixed non-uniform workload case, cloud services, database applications, and inference tasks assign uneven utilization across servers~\cite{narayanan2021efficient}.
These workload patterns directly affect power-system modeling.
The same aggregate IT demand can shift individual PSUs to different operating points and change their conversion losses.
However, existing studies do not translate these utilization patterns into a steady-state load model for power-flow analysis~\cite{khan2012workload,unuvar2014cloud,dong2023agent}.
We address this gap by developing an equivalent-circuit steady-state data center model.
Our model retains server-level utilization heterogeneity and captures load-dependent PSU conversion losses.

\noindent \textbf{Research contribution:} To address the limitations of existing aggregate models, this paper is the first to develop an equivalent-circuit steady-state data center model that captures how server loading affects PSU conversion losses and grid-side demand in power-flow studies, with the following contributions:

\begin{itemize}[leftmargin=*, nosep]
\item \textbf{Comprehensive equivalent-circuit model:} We develop a steady-state ECM that represents data-center IT, cooling, and auxiliary loads.
This model aggregates directly into an IV formulation for power flow.
\item \textbf{Loss-aware IT load characterization:} We explicitly model the IT load by combining utilization-dependent server power with ECMs of the PFC and LLC converter stages.
\item \textbf{Workload-to-grid coupling:} We incorporate heterogeneous server loading to quantify how different computational workload allocations alter internal converter losses.
This approach links server-level utilization directly to grid-side demand and transmission line stress.
\end{itemize}

We apply the proposed ECM to standard test systems and show that fixed-efficiency constant-PQ models can underestimate grid stress. In comparison with the constant-PQ model, the ECM predicts that the most stressed line exceeds its thermal limit in about 30\% of Monte Carlo samples. The results also show that homogeneous server utilization overstates line-loading variability by 17\%–46\% relative to heterogeneous server utilization.

\section{Equivalent Circuit Model of Data Center}
We develop the data center model at the component level using an equivalent-circuit formulation. 
The loads are connected to the UPS, but internally, in normal UPS operation mode, the loads are connected directly to the grid~\cite{jimenez2025data}. 
With a fully charged battery, we neglected the UPS conversion losses. 
Section~\ref{sec:IT_load_indi} models the IT load from server utilization and PSU conversion losses; Section~\ref{sec:Cooling} represents cooling with an induction-motor equivalent circuit, and Section~\ref{sec:Aux} models auxiliary demand as a constant impedance. Section~\ref{sec:full_load} aggregates these components into the total data center load for power-flow analysis.

\subsection{IT load model}\label{sec:IT_load_indi}
The IT load comprises server power demand and power supply conversion losses. We aggregate individual server-level IT loads to model the total cluster-level workload.

\subsubsection{Server load}

We represent the power consumption of a server with an affine function of the utilization factor~\cite{qureshi2009cutting}.
It models the power consumption of the server $j$ in cluster $i$ as
\begin{equation} \label{eq:ServerPower}
    P^{\mathrm{ser}}_{i,j} = P^{\mathrm{idle}}_{i,j} + \left(P^{\mathrm{max}}_{i,j} - P^{\mathrm{idle}}_{i,j}\right)\uti{i,j},
\end{equation}
where $u_{i,j}\in[0,1]$ denotes the utilization factor of the server $j$ in cluster $i$, and $P^{\mathrm{idle}}_{i,j}$ and $P^{\mathrm{max}}_{i,j}$ denotes its idle and maximum power consumption, respectively. We set $P^{\mathrm{idle}}_{i,j}=0.5P^{\mathrm{max}}_{i,j}$ the following~\cite{qureshi2009cutting}. Equation~\eqref{eq:ServerPower} maps server utilization to server-side power demand.
The utilization varies depending on the cluster workload. 
A homogeneous workload, such as LLM training, results in a homogeneous utilization factor within the cluster servers.
Agentic application, on the other hand, results in a heterogeneous utilization factor within servers of a cluster.

\subsubsection{Power supply units (PSU)}

At the rack level, PSUs convert AC input power and supply regulated DC power to the server loads.
We represent each server with one equivalent PSU that captures the aggregate behavior of its power-conversion interface.
Thus, the server $j$ in the cluster $i$ draws server-side DC power $P^{\mathrm{ser}}_{i,j}$, and its corresponding PSU supplies this demand while introducing conversion losses.
We model a PSU comprising a PFC converter and an isolated DC-DC LLC converter~\cite{InfineonAN145544}.
We construct the PSU equivalent-circuit formulation in two stages. The PFC stage maps the rack-side AC supply to the intermediate DC link. The LLC stage maps the DC link to the regulated server-side DC output.

\begin{figure*}[t!]
\includegraphics[width=\linewidth]{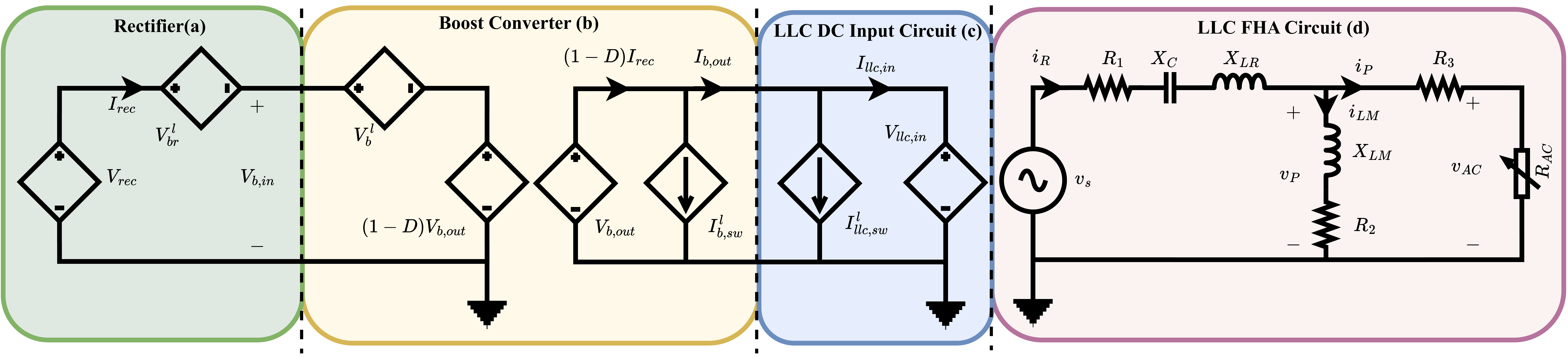}
\caption{Schematic and ECM of PFC + LLC, which includes a rectifier and the single switch boost converter, LLC DC input ECM, and LLC FHA AC ECM}
\label{fig:PSU_Full_ECF}
\end{figure*}

\paragraph{PFC converter ECM:}
Following~\cite{badmus2026two}, we represent the PFC stage using an equivalent circuit model (ECM) composed of a full-bridge diode. To capture load-dependent PSU losses in steady-state analysis, the dominant conduction and switching losses are represented in the ECM.
For the rectifier, the average rectified voltage and current are related to the sinusoidal input voltage and current by~\eqref{eq:VrecIrec}.
\begin{subequations}\label{eq:VrecIrec}
    \begin{align}
        \Vrec &= \frac{2}{\pi}|\VPFCin|\label{eq:Vrec}\\
        \Irec &= \frac{2}{\pi}|\IPFCin|\label{eq:Irec}
    \end{align}
\end{subequations}
Here, $|\VPFCin|$ and $|\IPFCin|$ denote the peak magnitudes of the sinusoidal input voltage and current, and $\Vrec$, $\Irec$ denote the lossless output voltage and current of the rectifier. The full-bridge rectifier uses four nominally identical diodes ($d_{1}$-$d_{4}$), each represented by the same forward voltage drop $\VF{0}$ and parasitic resistance $\RF{}$. 
During each conduction interval, the bridge current flows through two diodes.
We represent the bridge conduction effect by the equivalent voltage drop~\eqref{eq:Vbrloss}.
\begin{equation}\label{eq:Vbrloss}
        \Vbrloss = 2\VF{0} + 2\Irec\RF{}
\end{equation}
The loss, modeled as controlled voltage source, is the voltage difference between the lossless rectifier output voltage and the booster input voltage, as Fig.~\ref{fig:PSU_Full_ECF}(a) shows:
\begin{equation}
    \Vrec-\Vbrloss = \Vbin
\end{equation}
For the boost converter, we consider conduction loss and switching loss as the dominant loss mechanisms.
During the switch-on interval, the current flows through the inductor $L$ and switch $S$. 
During the switch-off interval, the current flows through the inductor $L$ (acts as a short during steady state) and diode $d$. 
Following the same equivalent-voltage-drop representation used in~\eqref{eq:Vbrloss}, we aggregate the average conduction effect of the boost stage as:
\begin{multline}\label{eq:Vbloss}
    \Vbloss = D\left[\VTl + \Irec \left( \RT + R_{\mathrm{Lb}} \right)\right] \\
    + (1-D)\left[\VDl + \Irec \left(\RD + R_{\mathrm{Lb}}\right)\right]
\end{multline}
where $D$ denotes the duty ratio of switch $S$, $\VTl$ and $\VDl$ denote the on-state voltage drops of the switch and boost diode, and $\RT$, $\RD$, and $R_{\mathrm{Lb}}$ denote the parasitic resistances of the switch, diode, and inductor, respectively.
In the steady-state model, $D$ is adjusted to regulate the boost output voltage, denoted by $\Vbout{}$ in Fig.~\ref{fig:PSU_Full_ECF}(c). Switching loss results from the overlap between the switch voltage and current during transistor transitions.
The corresponding switching-loss power is approximated as~\eqref{eq:PbSwLoss}.
\begin{equation}\label{eq:PbSwLoss}
    P^\mathrm{l}_{\mathrm{b,sw}}
    =
    \frac{1}{2}\Vbout{}\Irec(\ton+\toff)\fsw{pfc}
\end{equation}
The $\Vbout{}$ denotes the booster output voltage.
$\fsw{pfc}$ denotes the switching frequency of switch $S$ of PFC.
Following~\cite{badmus2026two}, this loss is represented as an equivalent current drawn from the boost output node, as given by~\eqref{eq:Ibswloss}.
\begin{equation}\label{eq:Ibswloss}
    \Ibswloss
    =
    \frac{P^\mathrm{l}_{\mathrm{b,sw}}}{\Vbout{}}
    =
    \frac{1}{2}\fsw{pfc}(\ton+\toff)\Irec
\end{equation}

\begin{figure*}[!t]
\centering
\includegraphics[width=1\linewidth]{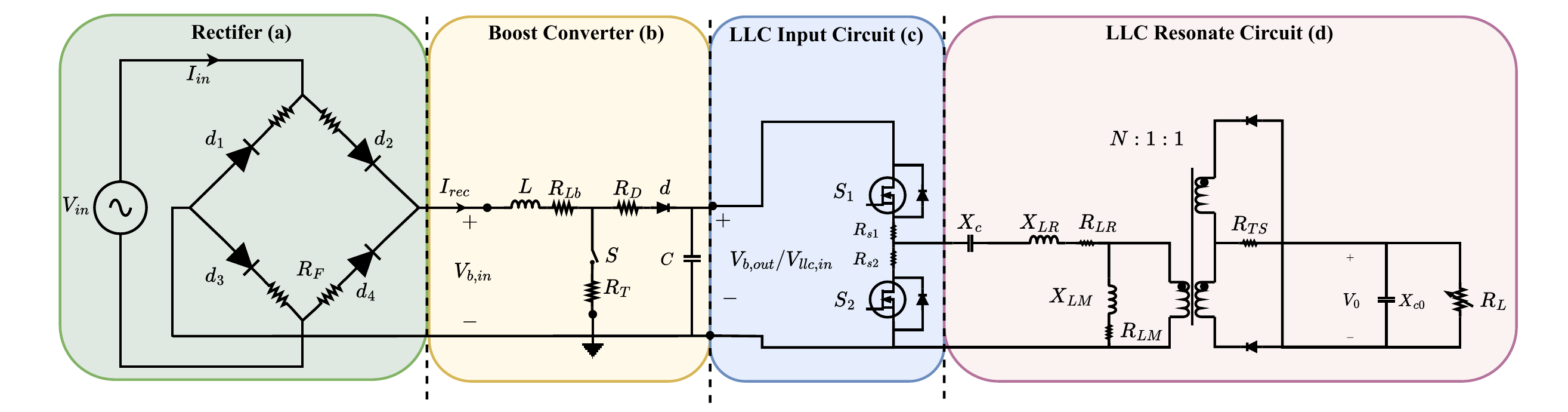}
\caption{Reference PSU architecture used in the proposed model, including the diode-bridge rectifier, boost PFC stage, LLC input-side circuit, and LLC resonant-tank circuit.}
\label{fig:PFC_LLC_main}
\end{figure*}

\noindent Figure~\ref{fig:PSU_Full_ECF}(a) and Figure~\ref{fig:PSU_Full_ECF}(b) show the equivalent circuit representations of the PFC converter.

\paragraph{LLC converter ECM:}
The PFC stage supplies the downstream LLC DC-DC converter through the intermediate DC link.
Following~\cite{yoo2022steady}, we decompose the LLC converter into an input-side circuit and a resonant-tank circuit.
Figure~\ref{fig:PFC_LLC_main}(c) and Figure~\ref{fig:PFC_LLC_main}(d) illustrate these circuits.
We construct two equivalent circuits based on this decomposition.
Figure~\ref{fig:PSU_Full_ECF}(c) shows an equivalent DC input circuit for the input side.
Figure~\ref{fig:PSU_Full_ECF}(d) shows a first-harmonic-approximation (FHA) equivalent circuit for the resonant tank~\cite{yoo2022steady}.
The FHA replaces the half-bridge square-wave excitation with its fundamental sinusoidal component.
This allows us to solve the resonant tank in the phasor domain at the switching frequency. 
The resulting resonant-current phasor provides the current input for the switching-loss model.

We represent the switching-loss effect of the two primary-side LLC switches $S_1$ and $S_2$ under switching frequency of $\fsw{llc}$  using an equivalent current-controlled current source. 
Under normal half-bridge operation, only one switch conducts at a time. 
However, each switch experiences one turn-off event per switching period. 
Because the Infineon reference design applies zero-voltage switching (ZVS)~\cite{InfineonAN145544}, we neglect the turn-on overlap loss and retain only the turn-off loss.
Let $\mathbf{I}_R$ denote the RMS FHA phasor of the resonant current. 
The corresponding time-domain resonant current is
\begin{equation}
    i_\mathrm{R}(t)=\sqrt{2}|\mathbf{I}_R|\cos(\omega_\mathrm{sw} t+\angle \mathbf{I}_\mathrm{R})
\end{equation}
Since the steady-state ECM does not explicitly track the switching instant, we approximate the turn-off current magnitude by the peak FHA resonant current,
\begin{equation}
    |i_{R,\mathrm{off}}| \approx \sqrt{2}|\mathbf{I}_\mathrm{R}|
\end{equation}
Therefore, the turn-off loss of the two primary-side switches is approximated as
\begin{equation}\label{eq:Pllcswloss}
    P^\mathrm{l}_{\mathrm{llc,sw}}= V_{\mathrm{llc,in}}\fsw{llc}\toff |i_{R,\mathrm{off}}|
\end{equation}
where $V_{\mathrm{llc,in}}$ denotes the input voltage to LLC.
Representing this loss as an equivalent input-side current source gives~\eqref{eq:Illcswloss}.
\begin{equation}\label{eq:Illcswloss}
    \Illcswloss = \fsw{llc}\toff \sqrt{2}|\mathbf{I}_\mathrm{R}|
\end{equation}
\noindent The current source $\Illcswloss$ represents the aggregate input-side current drawn by the LLC switching losses.
The detailed derivation of the FHA resonant-tank circuit with loss resistances follows~\cite{yoo2022steady} and is omitted for brevity. The LLC switching frequency is determined from the voltage-gain constraint.
We assume LLC delivers a constant voltage output, noted as $V_0$ in Fig.~\ref{fig:PFC_LLC_main}(d).
As PFC produces a constant voltage input to the LLC, and  LLC also delivers a constant voltage output, the voltage gain ($M$) of this LLC is a constant~\eqref{eq:VoltageGain}.
\begin{equation}\label{eq:VoltageGain}
    M = \frac{2NV_0}{V_\mathrm{llc,in}}
\end{equation}
In this work, we do not adopt the assumption of the original work where $R_1 = R_2 = R_3$\,. 
We retain the individual loss resistances as we benchmark against real equipment where $R_1 = R_\mathrm{s1/s2}+R_\mathrm{LR}$ (assume $S1$ and $S2$ identical),$R_\mathrm{2} = R_\mathrm{LM}$, and $R_\mathrm{3} \simeq \frac{8N^2}{\pi^2}R_\mathrm{TS}$ as it is the primary-side equivalent of $R_\mathrm{TS}$. 
Then we represent the LLC voltage gain using the squared form of the general FHA gain expression~\eqref{eq:llc_gain_constraint}.
\begin{equation}\label{eq:llc_gain_constraint}
\resizebox{0.95\linewidth}{!}{$
M
=
\frac{
\sqrt{
\left(\dfrac{R_2}{Z_0}\right)^2
+
\omega_\mathrm{n}^{2}K_\mathrm{L}^{2}
}
}{
\sqrt{
K_\mathrm{L}^2
\left[
\dfrac{R_\mathrm{K}^{\ast}}{Z_0^2K_\mathrm{L}}
+1-\omega_\mathrm{n}^{2}
\right]^2
Q^2
+
\omega_{n}^{2}
\left[
A+BK_\mathrm{L}-\dfrac{A}{\omega_\mathrm{n}^{2}}
\right]^2
}
}
$}
\end{equation}
Here we define the following terms for a compact representation: 
$A = 1+\frac{R_2+R_3}{R_\mathrm{AC}}$, 
$B = 1+\frac{R_1+R_3}{R_\mathrm{AC}}$, 
$R_K^{\ast} = R_1R_2+R_2R_3+R_3R_1+R_1R_\mathrm{AC}+R_2R_\mathrm{AC}$, 
$K_\mathrm{L}=\frac{L_\mathrm{M}}{L_\mathrm{R}}$, 
$Z_0=\sqrt{\frac{L_\mathrm{R}}{C_\mathrm{R}}}$, 
$\omega_\mathrm{n}=\frac{\omega_\mathrm{sw}}{\omega_0}=\frac{f_\mathrm{sw}^{\mathrm{llc}}}{F_\mathrm{R}}$, and 
$Q=\frac{Z_0}{R_\mathrm{AC}}$.

\begin{figure}
    \centering
    \includegraphics[width=\columnwidth]{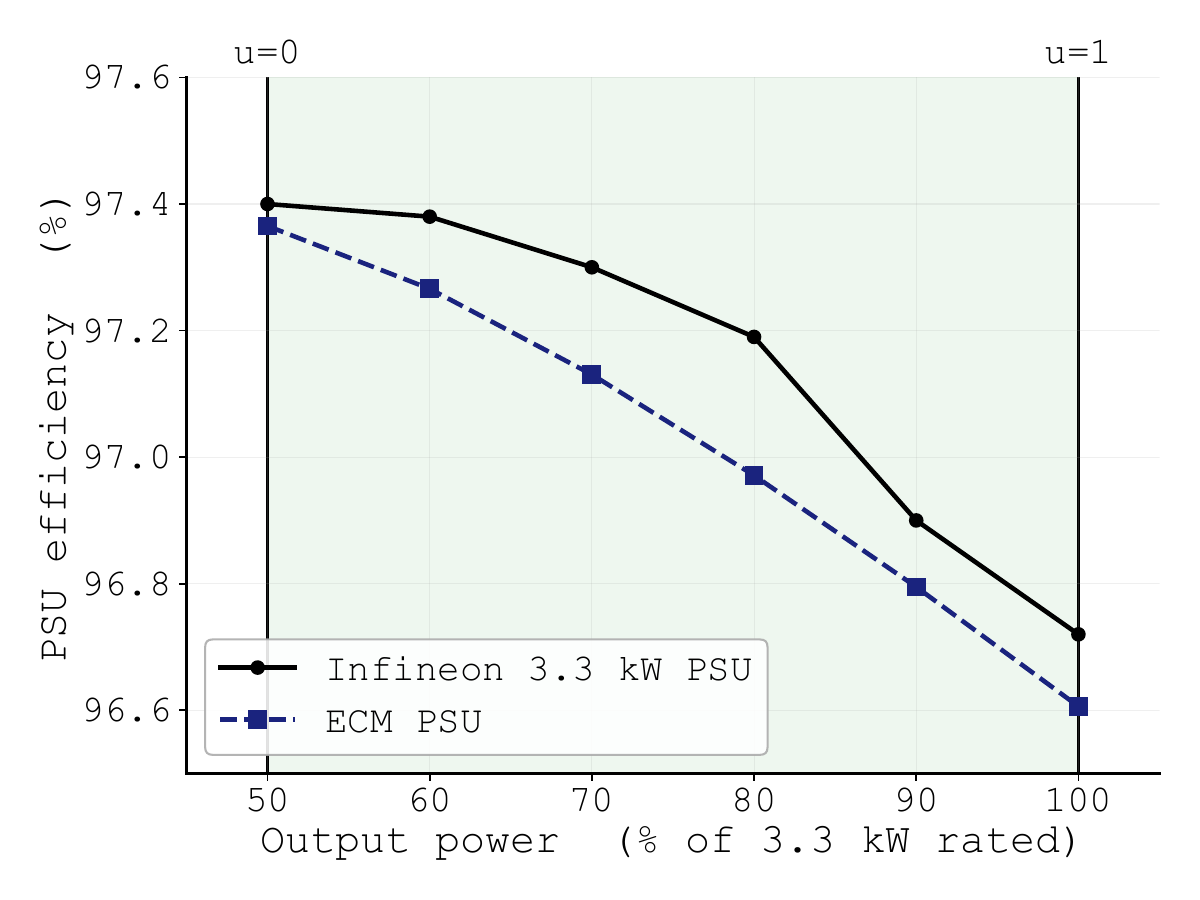}
    \caption{Comparison of manufacturer datasheet efficiency and the proposed PFC+LLC ECM efficiency.}
    \label{fig:psu_validation}
\end{figure}

In equation~\eqref{eq:llc_gain_constraint}, $M$ is a constant from~\eqref{eq:VoltageGain}. 
We solve for $\omega_\mathrm{n}$ across different values of $R_\mathrm{AC}$, where $R_\mathrm{AC} \simeq \frac{8N^2}{\pi^2}R_\mathrm{L}$ which is the primary-side equivalent of $R_\mathrm{L}$ shown in Fig.~\ref{fig:PSU_Full_ECF}(d).
$R_\mathrm{L}$ represents the server load resistance and changes with the server utilization.
Together, the PFC and LLC equivalent circuits form the PSU block used in the data-center steady-state model.
The PFC stage maps the rack-side AC input to the intermediate DC link, while the LLC stage maps the DC link to the regulated server-side DC output~\cite{yoo2022steady}.
This PSU representation allows the data center load to remain compatible with steady-state power-flow analysis while retaining internal converter variables, loss mechanisms, and DC-link operating conditions. Figure~\ref{fig:psu_validation} compares the efficiency predicted by the proposed PFC+LLC ECM with the reported efficiency of the Infineon 3.3-kW data center PSU over 50-100\% of rated output power~\cite{InfineonAN145544}. We obtain most ECM parameters directly from the datasheet and calibrate the remaining parameters against the reported efficiency curve. The resulting model closely reproduces the efficiency trend across the operating range.

\subsubsection{Cluster-level power aggregation of IT load}

We aggregate the server-level power demand and PSU losses to obtain the total demand of each of the $N^{\mathrm{clus}}$ clusters. For cluster $i$, we compute the active power by summing the server power and PSU losses across all servers $N^{s}_{i}$. Since the PSU operates at unity power factor~\cite{InfineonAN145544}, we set the cluster reactive power to zero.

\begin{equation}
P^{\mathrm{clus}}_{i}
=
\sum_{j=1}^{N^{s}_{i}}
\bigl(P^{\mathrm{ser}}_{i,j}
+
P^{\mathrm{PSU,loss}}_{i,j}\bigr),
\quad
Q^{\mathrm{clus}}_{i}=0 .
\label{eq:cluster_power}
\end{equation}

\subsection{Cooling system modeling}\label{sec:Cooling}

Cooling represents a major non-IT load in data centers and accounts for about 30\% of total power consumption in standalone facilities~\cite{jimenez2025data}. Cooling systems typically use motor-driven equipment, including compressors, pumps, fans, and chillers, which draw both active and reactive power from the AC distribution system. We therefore represent the cooling load at a fixed operating point, independent of IT loading, using the squirrel-cage induction-motor steady-state ECM presented in~\cite{pandey2016unified} which is compatible with the proposed circuit-based data center model. Since the cooling model serves as a standard supporting component rather than the main contribution of this work, we provide the full algebraic ECM equations in the accompanying open-source implementation~\cite{pandey2016unified}.

\subsection{Miscellaneous auxiliary load}\label{sec:Aux}
The miscellaneous load represents auxiliary data-center consumption not explicitly included in the IT or cooling-load models, including lighting, monitoring equipment, controls, networking support equipment, security systems, and other facility services. In general, these loads can be represented using a ZIP load model. However, because they are secondary to the proposed IT-load and PSU-loss models, this work approximates their aggregate effect as a constant-impedance load at the data center AC bus.

\subsection{Data center-level workload}\label{sec:full_load}

We obtain the full data-center demand by combining the server load from all clusters with the cooling and miscellaneous facility loads in~\eqref{eq:dc_power}. The active power includes server demand, PSU losses, cooling power, and miscellaneous power, while the reactive power comes from the cooling and miscellaneous loads.

\begin{subequations}
\label{eq:dc_power}
\begin{align}
P_{\mathrm{DC}}
&=
\sum_{i=1}^{N^{\mathrm{clus}}} P^{\mathrm{clus}}_{i}
+
P^{\mathrm{Cooling}}
+
P^{\mathrm{Misc}},
\label{eq:dc_power_P}
\\
Q_{\mathrm{DC}}
&=
Q^{\mathrm{Cooling}}
+
Q^{\mathrm{Misc}} .
\label{eq:dc_power_Q}
\end{align}
\end{subequations}

\section{Modeling Server Utilization}\label{sec:class_IT}
We model server utilization using two statistical methods: the Beta distribution and the copula method~\cite{nelsen2006introduction}. Each method serves a different purpose, and the workload type determines which one applies. The Beta distribution models a single utilization variable bounded between 0 and 1. Its shape parameters $\alpha$ and $\beta$ control the mean and spread, and prior studies use it to represent server and cluster utilization in data center workloads~\cite{unuvar2014cloud, dong2023agent}.  We use the Beta distribution directly when a single utilization value applies to an entire cluster.
A copula extends this representation to multiple servers by modeling their joint utilization. 
It assigns each server a Beta distributed utilization while controlling the correlation among servers in the same cluster through a single parameter $\rho_i$. 
We use a copula when servers operate at different but correlated utilization levels. 
The workload type determines the representation. 
Homogeneous workloads, such as synchronized parallel LLM training, distribute computation evenly and produce similar server utilization across a cluster~\cite{narayanan2021efficient} and we sample a scenario with Beta distribution. 
Heterogeneous workloads, such as cloud services and agentic AI applications, assign uneven but correlated server utilization due to different hardware requirements and execution paths~\cite{raj2026understandinganalyzingoptimizingagentic, khan2012workload}, and we sample a scenario with a copula model.

\subsection{Homogeneous workload allocation}\label{homo_explain}
Under homogeneous workload allocation, all servers in the same cluster operate at a uniform utilization factor. 

We sample the utilization factor of the server $j$ in the cluster $i$ from a Beta distribution
\begin{equation}\label{eq:ClusterUtilBeta}
u_{i,j} \sim \mathrm{Beta}(\alpha,\beta),
\end{equation}
where, $\alpha$ and $\beta$ can be selected from prior workload studies or fitted to measured cluster-utilization data. 
Under homogeneous allocation, every server in the cluster $i$ shares the same sampled value. We denote this shared value by $u_i$ and assign it homogeneously to every server:

\begin{equation}
\Userv{i,j} = \Uclus{i},
\quad \forall j \in \{1,\dots,N^{s}_{i}\},
\label{eq:HomoPerServer}
\end{equation}
Substituting~\eqref{eq:HomoPerServer} into the server-level model in~\eqref{eq:ServerPower} determines the server demand, and the cluster demand follows from the aggregation in~\eqref{eq:cluster_power}. If all servers and PSU modules in cluster $i$ are similarly loaded, the homogeneous cluster active power reduces to
\begin{equation}
    \footnotesize
    P^\mathrm{clus}_{i}
    =
    N^{s}_{i}
    \left[
    P^\mathrm{idle}_{i}
    +
    \left(P^\mathrm{max}_{i}-P^\mathrm{idle}_{i}\right)\Uclus{i}
    +
    P^{\mathrm{PSU,loss}}_{i}\!\left(\Uclus{i}\right)
    \right]
    \label{eq:HomoClusterPowerIdentical}
\end{equation}

\begin{figure}[t!]
\centering
\includegraphics[width=\columnwidth]{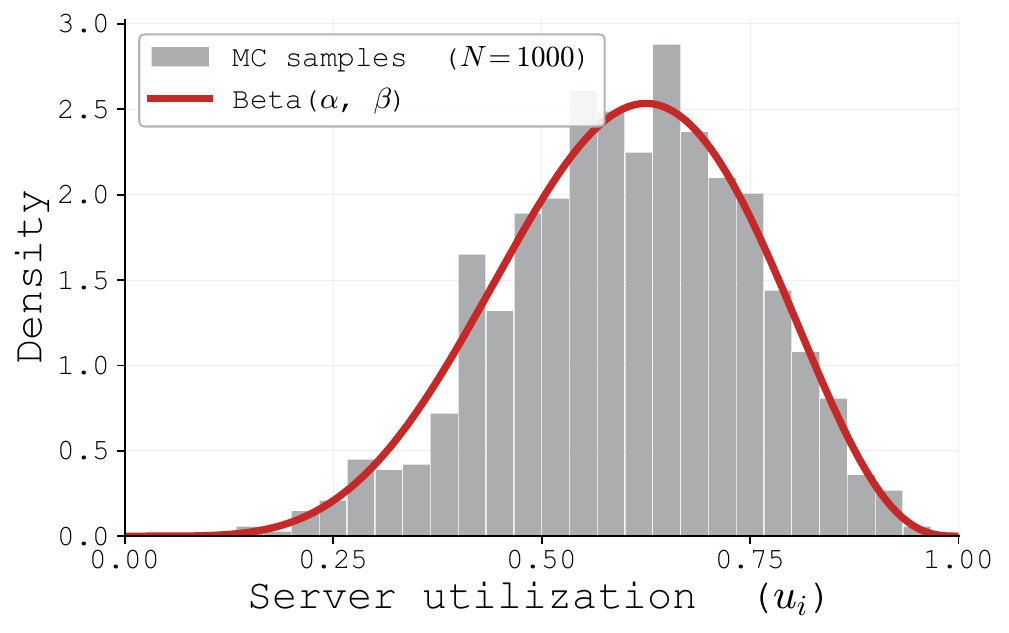}
\caption{Homogeneous server-utilization sampling for 1000 Monte Carlo trials. The histogram reports the sampled cluster-level utilization values, while the Beta density emphasizes moderate-to-high operating levels.}
\label{fig:server_utilization_homo}
\end{figure}

Figure~\ref{fig:server_utilization_homo} shows 1000 Monte Carlo samples $u_i$ drawn from the Beta distribution, whose probability mass concentrates around moderate-to-high utilization levels~\cite{shehabi20242024}.

\subsection{Heterogeneous workload allocation}\label{hetro_explain}
Under heterogeneous workload allocation, servers in the same cluster do not share a common utilization factor. Each server $j$ in the cluster $i$ draws an individual utilization factor $u_{i,j}$ with a $\text{Beta}(\alpha_i, \beta_i)$ marginal distribution, while a copula sets the correlation among servers in the same cluster~\cite{nelsen2006introduction}.
We let $g_i \sim \mathcal{N}(0,1)$ denote a common random factor for cluster $i$, and let $\epsilon_{i,j} \sim \mathcal{N}(0,1)$ denote the server-specific random factor for server $j$. 
For a prescribed correlation coefficient $\rho_i$, the latent normal variable of the server $j$ is given by~\eqref{eq:HetroLatentNormal}.
\begin{equation}\label{eq:HetroLatentNormal}
z_{i,j}
    =
    \sqrt{\rho_i} g_i
    +
    \sqrt{1-\rho_i}\epsilon_{i,j},
    \quad \forall j \in \{1,\dots,N^{s}_{i}\}
\end{equation}
We map $z_{i,j}$ through the standard normal CDF and the inverse Beta CDF to obtain the server utilization,
\begin{equation}\label{eq:HetroRawServerUtil}
    \Userv{i}{j}
    =
    F_{\mathrm{B}}^{-1}
    \left(
    \Phi(z_{i,j});\alpha_i,\beta_i
    \right),
\end{equation}
where $\Phi(\cdot)$ is the standard normal cumulative distribution function and $F_{\mathrm{B}}^{-1}(\cdot;\alpha_i,\beta_i)$ is the inverse cumulative distribution function of the Beta distribution. 
\begin{figure}[t!]
    \centering
    \includegraphics[width=\columnwidth]{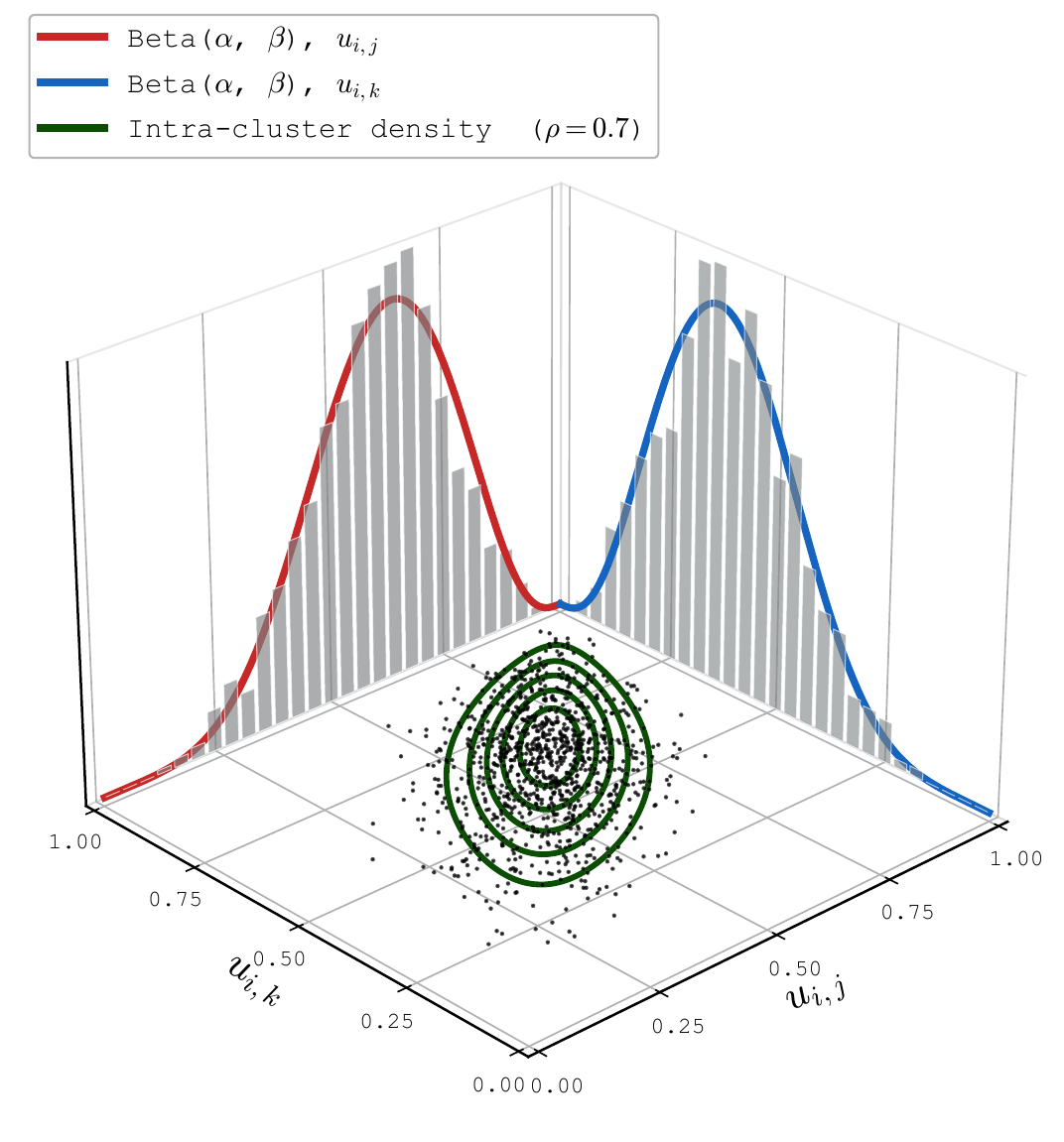}
    \caption{Heterogeneous server-utilization sampling. The Beta marginals shape individual server utilization, while the joint Monte Carlo samples show the intra-cluster dependence controlled by $\rho$.}
    \label{fig:server_utilization_hetro}
\end{figure}
Figure~\ref{fig:server_utilization_hetro} illustrates this construction. The marginal Beta distributions describe the utilization behavior of individual servers, while the joint scatter and density contours show how the correlation parameter $\rho_i$ controls the dependence between servers. As $\rho_i$ increases, server utilizations move more closely together; as $\rho_i$ decreases, servers operate more independently. Thus, the model separates the shape of individual server utilization from the dependence structure among servers in the same cluster.
Now the cluster-level utilization factor is,
\begin{equation}\label{eq:ClusterUtilWeighted}
    \Uclus{i}
    =
    \frac{
    \sum_{j=1}^{N^{s}_i}
    \left(P^{\mathrm{max}}_{i,j}- P^{\mathrm{idle}}_{i,j}\right)u_{i,j}
    }{
    \sum_{j=1}^{N^{s}_i}
    \left(P^{\mathrm{max}}_{i,j}-P^{\mathrm{idle}}_{i,j}\right)
    },
\end{equation}
while if all servers are identical, then~\eqref{eq:ClusterUtilWeighted} reduce to
\begin{equation}\label{eq:ClusterUtilFromServers}
    \Uclus{i}
    =
    \frac{1}{N^{s}_i}
    \sum_{j=1}^{N^{s}_i}
    \Userv{i}{j}.
\end{equation}

\section{Experiment Setup and Results}
We evaluate the proposed ECM on the \textit{savnw} 23-bus case and the Texas 2k-bus high-load case~\cite{birchfield2016grid}. 
We first add an MV/LV transformer interface to the power-flow equations at each selected PQ bus, then replace the original load model with the proposed data-center ECM. 
For each data center, the utilization-based server model computes individual server IT demand. 
Without loss of generality, we represent each data center with a single cluster, while the proposed ECM can model multiple clusters through cluster-specific utilization, server populations, and PSU models.

For each experiment, we run a Monte Carlo simulation with 1000 samples, computing the PFC and LLC conversion losses for each server and solving the steady-state power flow equations at each sample. 
We use line loading relative to thermal limits as the comparison metric because PSU loss models change the grid-side data center demand and directly affect nearby transmission flows. 
The following sections present two studies: a comparison of the proposed data center ECM against the constant-PQ model with fixed efficiency, and a comparison of the homogeneous and heterogeneous utilization-based IT load models.

\subsection{Comparison against constant efficiency P/Q model}
\begin{figure*}[!t]
\centering
\begin{minipage}[t]{0.48\textwidth}
\centering
\includegraphics[width=\linewidth]{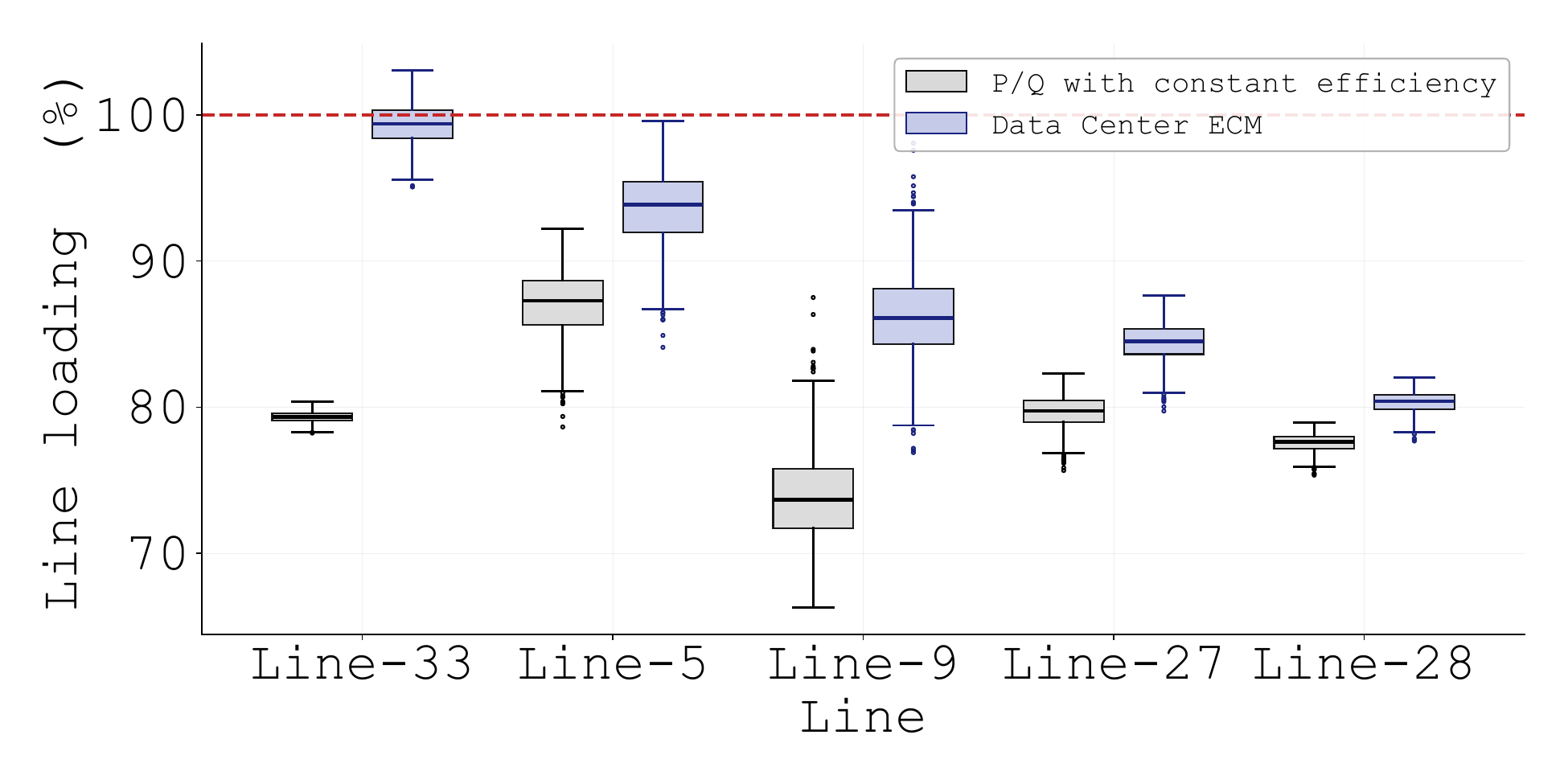}
\vspace{2pt}
(a) Line loading distributions for the five DC-adjacent transmission lines
\label{fig:volt_comp}
\end{minipage}
\hfill
\begin{minipage}[t]{0.48\textwidth}
\centering
\includegraphics[width=\linewidth]{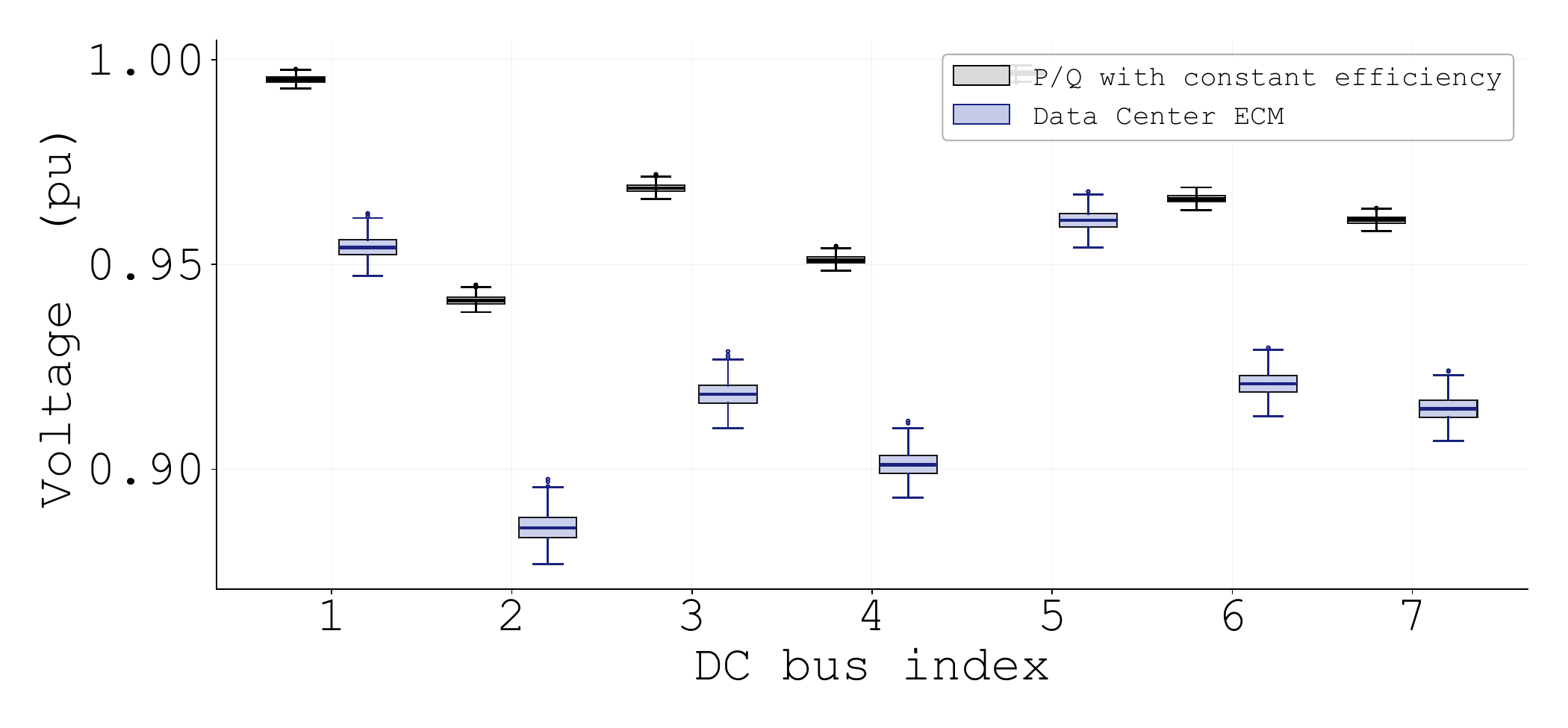}
\vspace{1pt}
(b) Transmission-bus voltage distributions at the seven data-center buses 
\label{fig:line_comp}
\end{minipage}
\caption{Comparison of P/Q model with constant efficiency and ECM data center over 1000 Monte Carlo samples.}
\label{fig:comparison_PQ_ECM}
\end{figure*}

We modify the \textit{savnw} 23-bus test network by replacing the $P/Q$ loads at seven selected buses with data center loads. We set the fixed PSU efficiency of IT load to 97\%, which corresponds to the maximum efficiency reported in the data sheet~\cite{InfineonAN145544}. 
For this comparison, we use the homogeneous loading case and assign a mean utilization of 0.8\footnote{Different utilization scenarios ~\cite{shehabi20242024,knittel2025flexible,guidi2024environmental} can be modeled by adjusting the Beta distribution parameters.} to represent a high-loading scenario~\cite{semianalysis2024, knittel2025flexible}. 

Figure~\ref{fig:comparison_PQ_ECM}(a) compares the line-loading and voltage distributions obtained from the constant-PQ model and the proposed ECM.
For Line 33, the proposed ECM predicts line-limit violations in about 30\% of the Monte Carlo samples, while the constant-PQ model keeps the line below its limit.
For the remaining lines, the ECM consistently shifts the loading distributions upward, indicating higher congestion risk.
The ECM also lowers the voltage distributions at the data center buses because it accounts for PSU losses and draws higher grid-side power as shown in~Fig.~\ref{fig:comparison_PQ_ECM}(b). 
These trends show that fixed-efficiency P/Q load models can underestimate the electrical stress caused by data-center loads.

\subsection{Comparison between homogeneous and heterogeneous workloads}
\begin{figure*}[!t]
\centering
\begin{minipage}[t]{0.48\textwidth}
\centering
\includegraphics[width=\linewidth]{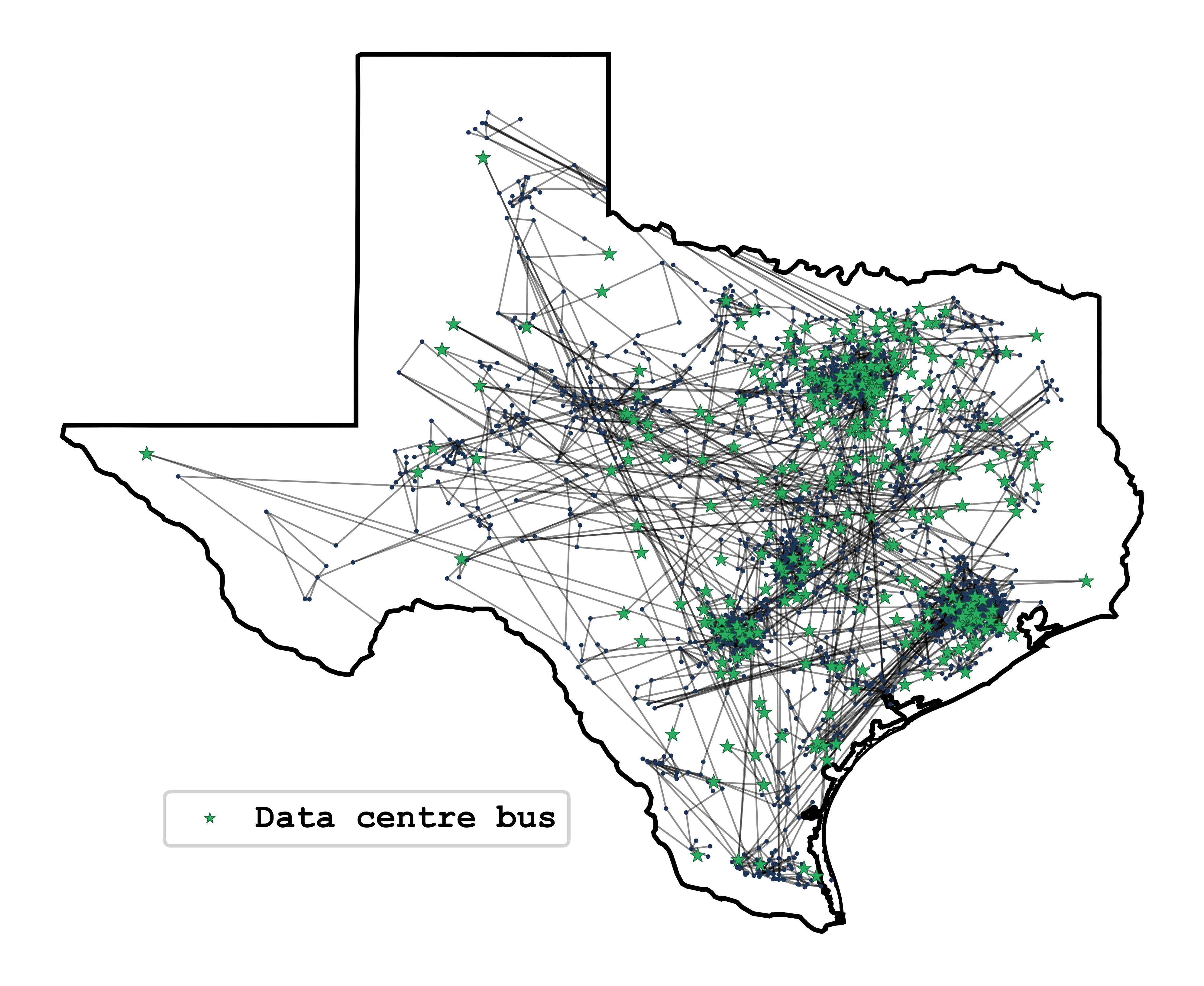}
\vspace{2pt}
(a) Texas 2k-bus system with selected data center locations.
\label{fig:texas_2k_dc}
\end{minipage}
\hfill
\begin{minipage}[t]{0.48\textwidth}
\centering
\includegraphics[width=\linewidth]{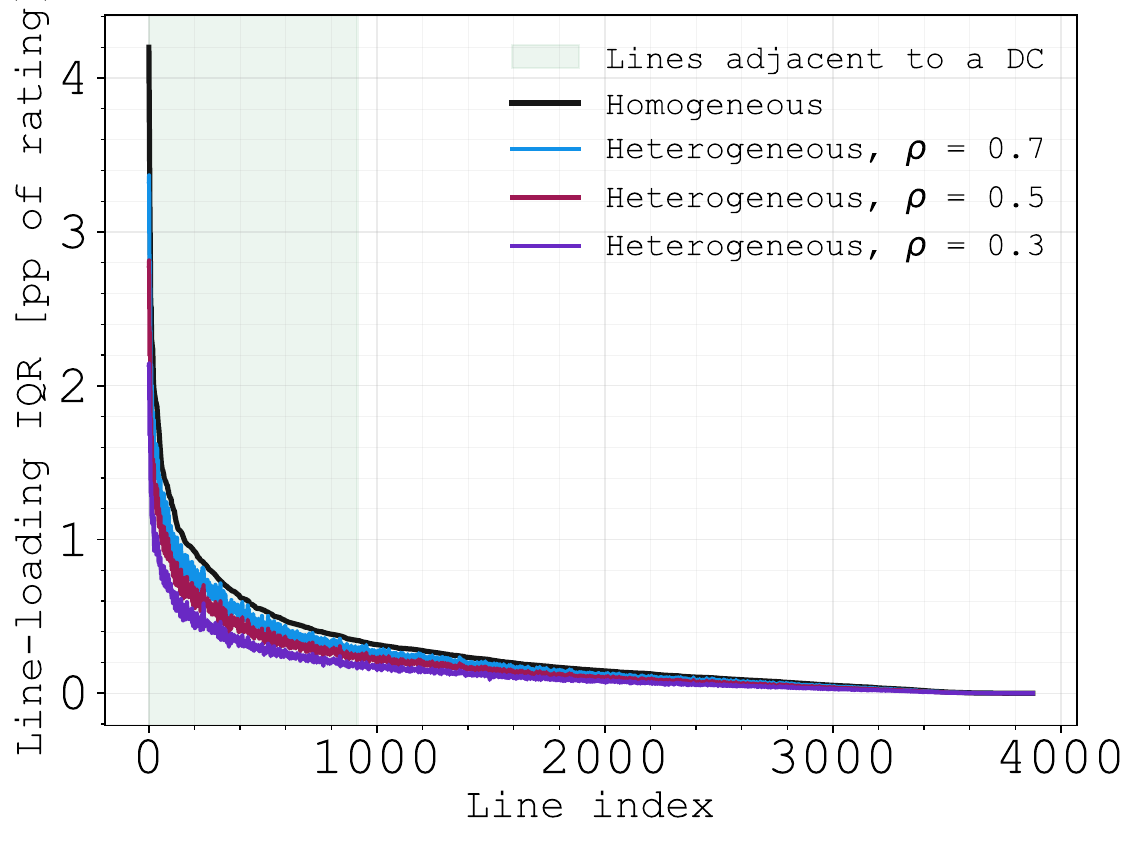}
\vspace{1pt}
(b) Line-loading inter-quartile range (IQR) per rated line, in percentage points (pp) of line rating.
\label{fig:iqr}
\end{minipage}
\caption{Texas 2k-bus data center placement and line-loading variability under the homogeneous and heterogeneous workload models.}
\label{fig:texas_iqr_side_by_side}
\end{figure*}

In this experiment, we compare two models of how server utilization varies within a data center cluster. 
In the homogeneous model, every server in a cluster draws the same utilization index for a given Monte Carlo sample, so the cluster behaves as a single aggregate load, as highlighted in section~\ref{homo_explain}.
In the heterogeneous model, each server draws an independent utilization index, linked through intra-cluster correlation $\rho$ as explained in detail in section~\ref{hetro_explain}. 
We evaluate both models on the Texas 2k-bus synthetic network~\cite{birchfield2016grid}, in which we replace the loads at 300 buses with data center loads at different locations across the grid, as shown in Fig.~\ref{fig:texas_iqr_side_by_side}(a). 
For both models, the utilization index follows a Beta($\alpha=6$, $\beta=4$) distribution with a mean of approximately 0.6, representing a moderate operating condition within the utilization range of 0.5-0.8 reported in recent data-center studies~\cite{shehabi20242024,knittel2025flexible,guidi2024environmental}.
As in the previous experiment, we run a Monte Carlo simulation with 1000 samples and compute the resulting line-loading distribution for each sample.

After solving the power flow for each Monte Carlo sample, we compute the inter-quartile range (IQR) of the line-loading distribution for every rated line. The IQR, defined as $\text{IQR}=Q_{75}-Q_{25}$, captures the spread of the middle 50\% of the loading samples. 
Figure~\ref{fig:texas_iqr_side_by_side}(b) shows the per-line IQR for all rated lines in the network, sorted in descending order based on the homogeneous-model IQR. 
The green shading identifies lines adjacent to data center buses. 
We quantify the network-wide effect by computing the mean IQR over all rated lines for each model\footnote{For each line \(i\), we compute
\(\mathrm{IQR}_i = Q_{75,i} - Q_{25,i}\) across the Monte Carlo samples. We report the resulting spread in percentage points (pp) of the line rating; for example, a change from \(40\%\) to \(43\%\) equals \(3\)~pp.}. 
The homogeneous model overstates the mean IQR by approximately 17.2\%, 30.5\%, and 46.5\% relative to the heterogeneous model at $\rho = 0.7$, $0.5$, and $0.3$, respectively. Lower intra-cluster correlation produces a narrower line-loading distribution, while higher correlation drives the heterogeneous model toward the homogeneous case.

This result shows that treating a data center cluster as a single aggregate load can overstate grid-side power variability and the resulting line-loading spread. In contrast, the heterogeneous model gives transmission operators a less conservative and more realistic estimate of loading variability and available transmission margin.
\section{Conclusion}
We develop a steady-state ECM that links computational workloads to grid-side data center demand by representing utilization-dependent IT loads, loss-aware PFC and LLC converter stages, cooling systems, and auxiliary loads. Simulation studies with Monte Carlo analysis on power-system test cases show that the ECM changes the predicted grid impact of data-center loads compared with fixed-efficiency constant-PQ models.
We highlight three key findings:
\begin{itemize}[leftmargin=*, nosep]
    \item \textbf{Equivalent circuit framework:} We build an aggregate equivalent circuit model for utilization-dependent server load with loss-aware converter models and induction motor. These models seamlessly fit into the IV power flow formulation without loss of generality.
    \item \textbf{Heterogeneous server loading:} The framework captures heterogeneous loading across clusters and servers. Assuming homogeneous cluster utilization overstates line-loading variability by $\approx$17\% to 46\% relative to the heterogeneous model. 
    \item \textbf{Prediction of grid stress:} The ECM predicts greater transmission-system stress than the baseline fixed-efficiency constant-PQ model. Under high loading, the ECM predicts line-limit violations in about 30\% of Monte Carlo samples, while the baseline model remains below the thermal limit.
\end{itemize}

These results can help utilities and planners identify potential grid constraints and determine appropriate interconnection requirements or additional facilities needed for stable operation.

\printbibliography
\end{document}